\documentclass[twocolumn,prd,showpacs,10pt]{revtex4}
\usepackage{graphicx}
\usepackage{dcolumn}
\usepackage{bm}

\newcommand{\square}{\kern1pt\vbox{\hrule height 1.2pt\hbox{\vrule
width 1.2pt\hskip 3pt
\vbox{\vskip 6pt}\hskip 3pt\vrule width 0.6pt}\hrule
height 0.6pt}\kern1pt}
\newcommand{\beq}{\begin{equation}}
\newcommand{\beqn}{\begin{eqnarray}}
\newcommand{\eeq}{\end{equation}}
\newcommand{\eeqn}{\end{eqnarray}}

\begin{document}

\title{Generalised cosmological scaling solutions}
\author{Edmund J.~Copeland$^1$, Seung-Joo Lee$^1$, James E. Lidsey$^2$
and Shuntaro Mizuno$^3$}
\address{$^1$ Department of Physics and Astronomy, University of Sussex,
Falmer, Brighton BN1 9QJ, United Kingdom}
\address{$^2$ Astronomy Unit, School of Mathematical Sciences, 
Queen Mary, University of London, Mile End Road,
London, E1 4NS, UK}
\address{$^3$ Department of Physics, Waseda University, Okubo 3-4-1,
Shinjuku, Tokyo 169-8555, Japan}

\date{\today}

\begin{abstract}
Motivated by the recent interest in cosmologies arising from 
energy density modifications to the Friedmann equation,  
we analyse the scaling behaviour for a broad class of these 
cosmologies comprised of scalar fields and background barotropic fluid 
sources. In particular, we determine the corresponding scalar 
field potentials which lead to attractor scaling solutions
in a wide class of braneworld and dark energy scenarios. 
We show how a number of recent proposals for modifying 
the Friedmann equation can be thought of as 
being dual to one another, and determine the conditions under 
which such dualities arise. 
\end{abstract}

\vskip 1pc 
\maketitle

%

\section{Introduction}

The Friedmann equation is one of the cornerstones of modern 
cosmology relating the expansion of the Universe to the 
total energy density within it. It forms the starting point 
for almost all investigations in cosmology. However, over 
the past few years, possible corrections to the Friedmann 
equation have been derived or proposed in a number of different 
contexts, generally inspired by braneworld investigations. These  
corrections are often of a form that involves the total 
energy density $\rho$, and are such that they tend to 
play a role early in the history of the Universe, fading away 
as we enter the late--time, post--nucleosynthesis era (although that is not 
always the case as we shall see). Up to now, the 
different models have been presented in the literature 
without any attempt to relate them. In this paper, by 
introducing a generalised form for the correction, we 
will provide a formalism which allows us to relate a 
large class of modified Friedmann cosmologies. 
Assuming the total energy density to be comprised of a 
canonical scalar field $\phi$ with potential $V(\phi)$, 
together with some form of barotropic fluid, we will demonstrate 
how the existence of scaling solutions determines the form of $V(\phi)$, 
and in doing so we will establish a direct relation between the form of 
the potential and the functional form of the modification to the 
Friedmann equation. 
Scaling (attractor) solutions in cosmology are very important because  
they allow one to understand the asymptotic behaviour of 
a particular cosmology and to determine whether such  
behaviour is stable or not. They have also been advocated by a 
number of authors as a way of establishing the behaviour 
of general scalar fields in a cosmological setting  both in the context of conventional 
Friedmann cosmologies 
\cite{Ratra_Peebles,Steinhardt,trac,CLW,ferreira,Liddle,moduli_stabilization,sahni,
delaMacorra:1999ff,Ng:2001hs} 
and in particular classes of modified Friedmann cosmologies 
\cite{Copeland:2000hn,Huey:2001ae,Sahni:2001qp,Huey:2001ah,Majumdar:2001mm,Nunes:2002wz,vandenhoogen,Lidsey:2003sj,Tsujikawa:2003zd,Sami:2004ic,Savchenko:2002mi,Calcagni:2004bh,Tsujikawa:2004dp}.

In section II we present the equations of motion arising 
out of the modified Friedmann equations and introduce variables 
which allow the scaling solutions to be determined. 
The general conditions for scaling behaviour are then 
established in Section III and we show that 
these can be written as a closed form relationship between the scalar 
field and the functional form of the modification to the 
Friedmann equation. In section IV we 
demonstrate the existence of duality symmetries between 
different scaling solutions, and determine   
the conditions which must be satisfied in terms of the 
modification to the Friedmann equation for such duality 
properties to be obtained. Section V then applies the 
results of the previous sections to a general class of models, 
evaluating the scaling potentials (and their duals), as well 
as the explicit evolution of the scale factor in the scaling 
regimes. In particular, we apply the technique to a number of recently 
investigated cosmologies of relevance both to braneworld and 
dark energy scenarios, including the 
Randall--Sundrum \cite{R_S}, 
Shtanov--Sahni \cite{Shtanov:2002mb} 
and Cardassian \cite{Freese:2002sq} models. 
The Dvali-Gabadadze-Porrati \cite{Dvali:2000hr} braneworld scenario 
is investigated in Section VI and 
we summarize our results in section VII. 
Throughout units are chosen such that $\hbar = c =1$.

\section{Equations of motion}

We consider spatially flat Friedmann--Robertson--Walker (FRW)
cosmologies such that the dynamics is determined by an 
effective Friedmann equation of the form
\beqn
\label{b_eq}
H^2 = \frac{8 \pi}{3 m_4^2} \rho L^2(\rho),
\eeqn
where $H \equiv \dot{a}/a$ is the Hubble parameter, $a$
is the scale factor, $\rho$ is the total energy density of the universe, 
a dot denotes differentiation with respect to cosmic 
time and $m_4$ is the four--dimensional Planck mass. 
Modifications to standard relativistic cosmology are 
parametrized by the correction
function $L(\rho )$ and this is assumed to be positive--definite 
without loss of generality. 

We will investigate models where the universe is   
sourced by a self--interacting scalar field $\phi$ with 
potential $V(\phi )$ together with a barotropic fluid 
with equation of state $P_{\gamma} = (\gamma - 1) \rho_{\gamma}$,
where $\gamma$ is the adiabatic index. The energy density and pressure of 
the
scalar field are given by $\rho_\phi = \dot{\phi}^2/2+V$ and 
$P_{\phi} = \dot{\phi}^2/2 -V$, respectively. As in conventional 
cosmologies, we assume that the energy--momenta of these matter 
fields is covariantly conserved and this implies that 
\beqn
\label{fluidconserve}
\dot{\rho}_{\gamma} &=& - 3\gamma H \rho_{\gamma},\\
\label{scalareom}
\ddot{\phi} &=& -3H\dot{\phi} - dV/d\phi.
\eeqn
Eqs. (\ref{b_eq})--(\ref{scalareom}) close the system that determines 
the cosmic dynamics.
 
In standard cosmology the stability of scaling solutions 
is analyzed by introducing the variables \cite{CLW}:
\begin{eqnarray}
\label{XCLW}
X_{\rm CLW} \equiv  \sqrt{\frac{4\pi}{3m_4^2}} \frac{\dot{\phi}}{H}
\nonumber
\\
\label{YCLW}
Y_{\rm CLW} \equiv \sqrt{\frac{8\pi}{3m_4^2}} \frac{\sqrt{V}}{H}
\end{eqnarray}
and rewriting the field equations as an autonomous 
system. Following \cite{Huey:2001ae,Huey:2001ah}, we define the new 
pair of variables: 
\beqn
\label{defX}
X &\equiv& \frac{\dot{\phi}}{\sqrt{2\rho}},\nonumber\\
\label{defY}
Y &\equiv& \frac{\sqrt{V}}{\sqrt{\rho}},
\eeqn
that are related to those of Eq. ({\ref{YCLW}) by 
\beqn
\label{relatedXY}
\frac{X}{X_{\rm CLW}} = \frac{Y}{Y_{\rm CLW}} =
\epsilon L,
\eeqn
where $\epsilon = \pm 1$ for expanding and contracting universes, 
respectively. In what follows, we consider expanding models unless 
otherwise stated. 

When expressed in terms of the new variables (\ref{defY}), 
the equations of motion (\ref{b_eq})--(\ref{scalareom}) can be written in 
the form: 
\beqn
\label{b_eq_x}
X' &=& -3X + \epsilon \lambda \sqrt{\frac{3}{2}} Y^2 
+ \frac{3}{2} X [2 X^2 + \gamma (1- X^2 - Y^2)],\nonumber\\
\\
\label{b_eq_y}
Y' &=& -\epsilon \lambda \sqrt{\frac{3}{2}} XY 
+ \frac{3}{2}Y[2X^2 + \gamma (1- X^2 - Y^2)],\nonumber\\
\\
\label{b_eq_l}
\lambda' &=& - \epsilon \sqrt{6} \lambda^2 (\Gamma -1) X\nonumber\\
&&+3 \lambda [2X^2 + \gamma (1-X^2-Y^2)] 
\rho \frac{d}{d \rho}[\ln L(\rho)],
\eeqn
where 
\beqn
\label{def_lambda}
\lambda &\equiv& -\frac{1}{L}\frac{m_4}{\sqrt{8 \pi}}
\frac{dV/d\phi}{V},\\
\label{def_Gamma}
\Gamma &\equiv& V \frac{d^2 V/ d\phi^2}{(dV / d\phi)^2},
\eeqn
and a prime denotes differentiation with respect to 
the logarithm of the scale factor, $N \equiv \ln a$.
As in Ref. \cite{Ng:2001hs},
Eqs. (\ref{def_lambda}) and (\ref{def_Gamma}) generalize 
the expressions introduced in 
Refs.~\cite{CLW,Huey:2001ah,delaMacorra:1999ff,Steinhardt}.
In particular, $\lambda$ is related to the parameter
$\lambda_{\rm CLW} \equiv -\frac{m_4}{\sqrt{8 \pi}}\frac{dV/d\phi}{V}$ 
introduced in Ref. \cite{CLW} such that 
\begin{equation}
\label{def_lclw}
\frac{\lambda}{\lambda_{\rm CLW}} = \frac{1}{L}   .
\end{equation} 
We refer to $\lambda$ as the `scaling parameter'.

The system of equations (\ref{b_eq_x})--(\ref{def_Gamma}) for 
the variables $X$, $Y$ and $\lambda$ do not appear to be closed due 
to the presence of the term involving $\Gamma$ in Eq. (\ref{b_eq_l}). 
However, it follows from Eqs. (\ref{defY}), (\ref{def_lambda}) and 
(\ref{def_Gamma}) that we have $\Gamma = \Gamma (\phi )$, 
$\phi = \phi (\rho , Y)$ and $\rho = \rho (\lambda , \phi )$ and 
hence that $\Gamma = \Gamma (\lambda , Y)$. This implies that the 
equations are indeed closed. 

The definition of the total energy density implies that 
the variables (\ref{defY}) satisfy the constraint equation
\beqn
X^2 + Y^2 +\frac{\rho_{\gamma}}{\rho} = 1
\eeqn
and, since the energy density of the barotropic fluid is 
semi--positive--definite, any cosmological model 
can be represented as a trajectory
in the $(X,Y)$--plane that is bounded within the unit circle, i.e., 
$\Omega_{\phi} \equiv \rho_\phi/\rho = X^2 + Y^2 \leq 1$. 
Furthermore, since $Y \geq 0$ by definition, it 
is sufficient to consider the evolution in the upper half of the disc.

Eqs.~(\ref{b_eq_x}-\ref{b_eq_l}) exhibit an important property. 
For the case where $\lambda$ is constant, they have an identical form to 
that of the plane--autonomous system of 
standard relativistic cosmology that is formulated in terms of the 
variables $\{ X_{\rm CLW}, Y_{\rm CLW}, \lambda_{\rm CLW} \}$.
This duality immediately implies that the system 
(\ref{b_eq_x}-\ref{b_eq_y}) admits an identical set of critical points 
to that of the standard scenario when these solutions are expressed in 
terms of the variables $\{ X ,Y, \lambda \}$. 

A further consequence of such a duality 
is that the stability of each fixed point 
solution can be determined directly 
from the stability analysis of Ref. \cite{CLW}. 
In total, there are five critical solutions to  
Eqs.~(\ref{b_eq_x}) and (\ref{b_eq_y}) where 
the variables $\{ X, Y, \lambda \} = \{ X_c, Y_c, \lambda_c \}$ are 
constants. Three of these represent the unstable solutions 
$(X_c=1,~Y_c=0)$,$(X_c=-1,~Y_c=0)$,
$(X_c=0,~Y_c=0)$ for all values of $\lambda$ and $\gamma$. 
The value of $\lambda$ determines the nature of the other two points. 
For $\lambda^2 > 3\gamma$, there exists an attractor solution 
\beqn
\label{scaling_sol}
X_c &=& \sqrt{\frac{3}{2}}\frac{\gamma}{\lambda},\nonumber\\
Y_c &=& \sqrt{\frac{3(2-\gamma) \gamma}{2\lambda^2}},
\eeqn
where the effective adiabatic index of the scalar field, defined by 
\begin{equation}
\label{defgammaphi}
\gamma_\phi \equiv \frac{2X^2}{X^2 + Y^2}  ,
\end{equation}
satisfies the condition $\gamma_{\phi} =\gamma$. For this 
late--time attractor solution, 
the relative contribution of the scalar field's energy density to the 
total energy density of the universe is constant, 
$\Omega_{\phi c} \equiv X^2 + Y^2 = 
3\gamma /\lambda^2$, and consequently, the energy densities 
of the scalar field and fluid redshift at the same rate as the universe 
expands. 

The fifth critical point arises if $\lambda^2 < 6$ and is given by 
\beqn
\label{p_l_sol}
X_c &=& \frac{\lambda}{\sqrt{6}},\nonumber\\
Y_c &=& \sqrt{1-\frac{\lambda^2}{6}} .
\eeqn
This corresponds to the case where the scalar field 
dominates the fluid $(\Omega_{\phi c} = 1)$ and has an 
effective adiabatic index $\gamma_\phi = \lambda^2 /3$. 
The solution is stable if $\gamma_\phi < \gamma$, i.e., 
$\lambda^2 < 3 \gamma$.

\section{General Conditions for Scaling Solutions}

Since the expressions (\ref{relatedXY}) and (\ref{def_lclw}) 
relating the standard and 
modified FRW cosmologies involve the correction function $L(\rho )$, 
the scalar field potential that gives rise to the 
fixed point attractor solutions 
(\ref{scaling_sol}) and (\ref{p_l_sol}) in a given generalized scenario  
will depend on the specific form of this function. In particular, 
the potential will differ from the purely exponential form
that leads to scaling solutions in the conventional FRW model.  
In this section we establish the correspondence between the 
modified Friedmann equation and the scaling potential. 

It can be shown by direct substitution 
that both sets of critical points (\ref{scaling_sol}) 
and (\ref{p_l_sol}) represent solutions to the field 
equations (\ref{b_eq_x})--(\ref{b_eq_l}) of the form 
$X' = Y' = \lambda'=0$ if the relation
\beqn
\label{dif_V_1}
\Gamma = 1+ \rho \frac{d}{d \rho} [\ln L(\rho)]
\eeqn
is satisfied. Since $\rho = V / Y_c^2$ for these 
solutions, Eq.~(\ref{dif_V_1}) may be written in the form
\beqn
\label{dif_V_3}
\rho\frac{d^2 \rho/d\phi^2}{(d\rho/d\phi)^2}-1-\rho \frac{d}{d\rho} 
[\ln L]=0,
\eeqn 
and multiplying Eq. (\ref{dif_V_3}) by $(d\rho/d\phi)/\rho$ then implies that
\beqn
\label{dif_V_4}
\frac{d \bigl(\ln (d\rho /d\phi)\bigr)}{d \phi} - 
\frac{d \bigl( \ln \rho \bigr)}{d \phi} -
\frac{d \bigl( \ln L \bigr)}{d \phi} = 0 .
\eeqn

Eq. (\ref{dif_V_4}) may be integrated twice to yield a 
necessary and sufficient condition on the scalar field potential 
if the solution is to represent a scaling 
solution for a given choice 
of correction function $L(\rho )$. We find the important result:
\beqn
\label{dif_V_2}
\int \frac{d\rho }{\rho L} = -\frac{\sqrt{8\pi}\lambda}{m_4}
\phi ,
\eeqn
where one of the integration constants has been set to zero without loss of 
generality by performing a linear shift in the value of the scalar field 
and the constant of proportionality on the right--hand side 
follows by requiring consistency with Eq.~(\ref{def_lambda}). 

It is also of interest to determine the evolution of the  
scale factor for a given class of scaling solutions. Since $X_c$ 
is a non--zero constant for these solutions,  
Eq. (\ref{defY}) implies that the scalar field is a monotonically 
varying function of proper time $(\dot{\phi} \ne 0)$. 
It is natural, therefore, to view the value of the field as 
the dynamical variable of the system and to express all 
time--dependent parameters in terms of this variable. 

In general, the scalar field Eq.~(\ref{scalareom}) 
can be expressed in the form
\begin{equation}
\label{rhoeom}
\dot{\rho}_{\phi} = - 3 H \dot{\phi}^2
\end{equation}
or, equivalently, as 
\begin{equation}
\label{rhoprime}
\frac{d \rho_{\phi}}{d \phi} = - 3 H \dot{\phi}  .
\end{equation}
It then follows from the definition of the Hubble parameter that 
\begin{equation}
\label{defH} 
3H^2 = - \frac{1}{a} \frac{da}{d \phi} \frac{d \rho_{\phi}}{d \phi}
\end{equation}
and substituting Eq. (\ref{defH}) into Eq. (\ref{b_eq}) 
implies that the Friedmann equation can be expressed in the form 
\begin{equation}
\label{friedmannprime}
\frac{da}{d\phi} \frac{d\rho}{d\phi} = - 
\frac{8\pi}{\Omega_{\phi c} m_4^2} a \rho L^2(\rho )  .
\end{equation}
Introducing a new variable 
\begin{equation}
\label{defb}
b (\phi ) = \exp \left[ \Omega_{\phi c} \int^{\rho} d \rho 
\frac{1}{\rho L^2(\rho )} \right]
\end{equation}
simplifies Eq. (\ref{friedmannprime}) to  
\begin{equation}
\label{friedmannsimpler}
\frac{da}{d\phi} \frac{db}{d\phi} = - \frac{8 \pi}{m_4^2} a b
\end{equation}
and the scale factor 
is then determined up to a single quadrature: 
\begin{equation}
\label{quadrature}
a(\phi ) = \exp \left[ - \frac{8\pi}{m_4^2} 
\int^{\phi} d \phi b(\phi ) \left( 
\frac{db}{d\phi} \right)^{-1} \right]  .
\end{equation}

Thus, for a given cosmological scenario characterized 
by a correction function $L(\rho )$, the potential (and equivalently 
the total energy density) yielding the scaling solution is 
determined by integrating Eq. (\ref{dif_V_2}). 
Integration of Eq. (\ref{defb}) then yields the dependence of 
$b(\phi )$ on the scalar field and the evolution of the scale 
factor follows after integration of Eq. (\ref{quadrature}). Finally, 
the time--dependence of the scale factor can in principle be deduced by 
integrating Eq. (\ref{rhoprime}),  
\begin{equation}
\label{tdep}
t =- \frac{\sqrt{24\pi}}{\Omega_{\phi c} m_4} \int^{\phi}  d \phi 
L (\phi) \rho^{1/2} (\phi)  \left( \frac{d \rho}{d \phi} \right)^{-1}  ,
\end{equation}
and inverting the result. 

In the following section, we employ the above formalism to establish
a link between different classes of scaling solutions that arise for 
various choices of the modification to the Friedmann equation. 

\section{Duality between Scaling Solutions}

A duality between different scaling
solutions can be established by noting that 
Eq. (\ref{friedmannsimpler}) is invariant under the 
simultaneous interchange 
\begin{equation}
\label{simultan}
a (\phi ) \rightarrow b^p (\phi ), \qquad b (\phi ) 
\rightarrow a^{1/p} (\phi ) , 
\end{equation}
where $p$ is an arbitrary constant. This symmetry 
implies that a given scaling solution may be employed as a seed to generate 
a new scaling cosmology for a different Friedmann equation 
and associated scalar field potential. To be specific, 
let us consider the scaling solution  
parametrized by the functions $\{ a_+ (\phi ) , b_+ (\phi ) , 
\rho_+ (\phi )\}$ that arises for a specific choice of
correction function $L_+ (\rho )$. 
We now denote the `dual' scaling solution as $\{ a_- (\phi ) , b_- (\phi ) , 
\rho_- (\phi )\}$  and assume an {\em ansatz} of the form 
\begin{equation}
\label{ansatz}
b_- (\phi) = [a_+ (\phi ) ]^{1/p}  .
\end{equation}
The new scale factor is then determined from Eq. (\ref{quadrature}): 
\begin{equation}
\label{dualscale}
a_- (\phi ) = \exp \left[ -\frac{8\pi p}{m_4^2}
\int d\phi a_+ (\phi) \left( \frac{d a_+}{d\phi} \right)^{-1} \right] .
\end{equation}
However, since the function $a_+(\phi)$ is itself a solution 
to the Friedmann equation (\ref{friedmannsimpler}), 
Eq. (\ref{dualscale}) simplifies after integration to 
\begin{equation}
\label{newscale}
a_- (\phi ) = [b_+(\phi )]^p
\end{equation}
modulo an arbitrary (constant) prefactor.

We may now determine the condition that the dual 
correction function $L_- (\rho)$
must satisfy for the solution (\ref{newscale}) to also 
represent a scaling solution that satisfies 
Eq. (\ref{dif_V_2}). If we assume {\em a priori} that 
the two solutions $a_{\pm} (\phi)$ represent 
scaling solutions characterized by $\lambda_{\pm}$, respectively, 
Eq. (\ref{dif_V_2}) implies that 
\begin{equation}
\label{same}
\frac{1}{\lambda_+ \rho_+ L_+}
\frac{d\rho_+}{d\phi} = 
\frac{1}{\lambda_- \rho_- L_-} \frac{d\rho_-}{d\phi} 
\end{equation}
It then follows, after substitution of  Eq. (\ref{friedmannprime}) 
into the right--hand side of Eq. (\ref{same}), that 
\begin{equation}
\label{step1}
\frac{1}{\rho_+ L_+} \frac{d\rho_+}{d\phi} = - 
\frac{8\pi \lambda_+}{\Omega_{\phi c}m_4^2 \lambda_-} 
\frac{L_- a_-}{(da_-/d\phi )	}
\end{equation}
and Eq. (\ref{friedmannsimpler}) 
then implies that 
\begin{equation}
\label{step2}
\frac{1}{\rho_+ L_+}\frac{d\rho_+}{d\phi} =
\frac{\lambda_+}{\lambda_-\Omega_{\phi c}}
\frac{L_-}{b_-}\frac{db_-}{d\phi}    .
\end{equation}

On the other hand, 
substituting the {\em ansatz} (\ref{ansatz}) into Eq. (\ref{step2})
and employing 
Eq. (\ref{friedmannprime}) for the positive--branch solution 
yields the condition 
\begin{equation}
\label{step4}
\frac{1}{\rho_+^2 L_+^2} \left( \frac{d\rho_+}{d \phi} \right)^2 = 
- \frac{8\pi}{\Omega^2_{\phi c}m_4^2} \frac{\lambda_+}{\lambda_- p} 
L_-L_+  .
\end{equation}
Consistency with Eq. (\ref{dif_V_2}) therefore implies 
that a necessary and sufficient condition for the dual cosmology 
$\{ a_- (\phi ), L_- (\phi ) \}$ to represent a scaling 
solution is that {\em the correction functions arising in the 
respective Friedmann equations must be proportional to
the inverse of each other 
when both are expressed as functions of the scalar field}: 
\begin{equation}
\label{necessary}
L_+ (\phi ) L_- (\phi ) = - p \lambda_+ \lambda_- \Omega^2_{\phi c} .
\end{equation}
It is interesting that   
the standard relativistic cosmology $(L=1)$ represents the self--dual model
when $p \Omega_{\phi c}^2 \lambda_+\lambda_- =-1$. 

Finally, we find after substituting Eq. (\ref{necessary}) into 
Eq. (\ref{same}) and employing Eq. (\ref{defb}) that 
the energy density of the dual scaling solution is given by 
\begin{equation}
\label{energydual}
\frac{1}{\rho_-} \frac{d\rho_-}{d\phi} = -p \Omega_{\phi c} \lambda_-^2
\frac{1}{b_+} \frac{db_+}{d \phi}  .
\end{equation}
The dual potential then follows immediately from 
Eqs. (\ref{defY}) and (\ref{defb}): 
\begin{equation}
\label{dualpot}
V_-(\phi ) = Y_{c}^2\exp \left[ -p \lambda_-^2 \Omega^2_{\phi c}
\int d\rho_+ \frac{1}{\rho_+L_+^2} \right]   .
\end{equation}

In the following section we employ the techniques 
developed above to determine scaling solutions (and their duals) 
in a number of different cosmological settings.  

\section{Unification of modified Friedmann cosmologies}

\subsection{A Generalized Class of Scaling Cosmologies}

A wide class of scenarios that have been considered recently predict 
deviations from the standard cosmology of the form 
\beqn
\label{L_form}
L (\rho ) = \sqrt{1+A\rho^\nu}, 
\eeqn
where $\nu$ is an arbitrary dimensionless constant
and $A$ is an arbitrary  constant with dimension $m^{-4\nu}$.
In this case, the form of the potential leading to scaling (fixed point) 
solutions is determined 
by integrating Eq. (\ref{dif_V_2}). It is found that 
\beqn
\label{g-pot.1}
V_+(\phi) = Y_c^2 A^{-1/\nu} {\rm cosech}^{2/\nu} \,
\left( -\frac{\lambda \nu }{2}\frac{\sqrt{8 \pi}}{m_4} 
\phi \right)  ,
\eeqn
if $A > 0$ and
\beqn
\label{g-pot.2}
V_-(\phi) = Y_c^2 |A|^{-1/\nu} {\rm sech}^{2/\nu} \,
\left( -\frac{\lambda \nu}{2}\frac{\sqrt{8 \pi}}{m_4} \phi 
\right)   ,
\eeqn
if $A < 0$.

Given the form of the scalar potential (\ref{g-pot.1}), 
the parametric solution for the case $A>0$ is determined by 
integrating Eqs. (\ref{defb}) and (\ref{quadrature}), respectively: 
\begin{eqnarray}
\rho_+ (\phi) =  A^{-1/\nu} {\rm cosech}^{2/\nu} 
\, \left( -\frac{\lambda \nu}{2}\frac{\sqrt{8 \pi}}{m_4} \phi 
\right) 
\nonumber \\
b_+ (\phi) = A^{-\Omega_{\phi c} /\nu} 
{\rm sech}^{2\Omega_{\phi c} /\nu} \, 
\left( -\frac{\lambda \nu}{2}\frac{\sqrt{8 \pi}}{m_4} \phi 
\right) 
\nonumber \\
a_+(\phi ) = \sinh^{2/(\lambda^2\Omega_{\phi c} \nu)} \, 
\left( -\frac{\lambda \nu}{2}\frac{\sqrt{8 \pi}}{m_4} \phi 
\right)   .
\label{plussolution}
\end{eqnarray}
The time dependence of the solution follows by substituting Eq. 
(\ref{plussolution}) into Eq. (\ref{L_form}) to yield 
the Friedmann correction function: 
\begin{equation}
\label{plusL}
L_+(\phi ) = {\rm cotanh} \, 
\left( -\frac{\lambda \nu}{2}\frac{\sqrt{8 \pi}}{m_4} \phi 
\right) 
\end{equation}
and then evaluating the integrand in Eq. (\ref{tdep}): 
\begin{equation}
\label{integrandplus}
t= - \frac{\sqrt{3 A^{1/\nu}}}{\lambda \Omega_{\phi c}} 
\int d \phi \, {\rm sinh}^{1/\nu} 
\left( -\frac{\lambda \nu}{2}\frac{\sqrt{8 \pi}}{m_4} \phi 
\right)  .
\end{equation}
The integral (\ref{integrandplus}) can be performed analytically 
for various choices of $\nu$, whereas 
the late--time behaviour can be analysed 
for arbitrary $\nu$. In particular, 
we find from Eq.~(\ref{integrandplus}) that the late--time 
limit corresponds to large $\phi$, and it therefore follows from 
Eq.~(\ref{plusL}) that $L \rightarrow 1$ 
as $t \rightarrow \infty$. This 
in turn is the limit corresponding to the case of 
an exponential potential. 

The corresponding scaling solution for $A<0$ driven by the 
potential (\ref{g-pot.2}) is deduced by applying the duality 
transformation (\ref{simultan}) to the solution 
(\ref{plussolution}) for a particular value of the constant 
$p$, where the scaling parameters are chosen to be equal, 
$\lambda_+= \lambda_- = \lambda$. 
For the case where $p= -1/(\lambda^2 \Omega_{\phi c}^2)$, 
the duality transformation (\ref{necessary}) implies that 
the dual correction function is given by 
\begin{equation}
\label{minusL}
L_-(\phi ) = {\rm \tanh} \,  
\left( -\frac{\lambda \nu}{2}\frac{\sqrt{8 \pi}}{m_4} \phi 
\right) , 
\end{equation}
whilst integrating Eq. (\ref{dualpot}) with the form 
for $\rho_+(\phi )$ given in Eq. (\ref{plussolution}) implies that 
the dual potential has precisely the form of  
Eq. (\ref{g-pot.2}). We may conclude, therefore, 
that the dual correction function is given by Eq. (\ref{L_form}) 
with $A<0$. In this sense, a model with $A>0$ and a specific 
value of $\nu$ is twinned with the model where the value of $\nu$ 
is the same but the sign of $A$ is changed. 
In general, the dual scale factor is deduced from 
Eqs. (\ref{newscale}) and (\ref{plussolution}): 
\begin{equation}
\label{aminussolution}
a_-(\phi ) = \cosh^{2/(\lambda^2\Omega_{\phi c} \nu)} \, 
\left( -\frac{\lambda \nu}{2}\frac{\sqrt{8 \pi}}{m_4} \phi 
\right) 
\end{equation}
and the time--dependence follows from 
Eq. (\ref{tdep}): 
\begin{equation}
\label{integrandminus} 
t= - \frac{\sqrt{3 |A|^{1/\nu}}}{\lambda \Omega_{\phi c}} 
\int d \phi \, {\rm cosh}^{1/\nu} 
\left( -\frac{\lambda \nu}{2}\frac{\sqrt{8 \pi}}{m_4} \phi 
\right)   .
\end{equation}

It is of course trivial to show that for this choice of $p$,  
the standard cosmology solution corresponding to the 
case $L=1$ reproduces the well known exponential potential for the 
scalar field \cite{Lucchin}. 

In the following subsections, we consider some of the specific models
that belong to the class of corrections given by Eq.~(\ref{L_form}). 

\subsection{Randall-Sundrum Type II braneworld cosmology}

The case $A=1/2\sigma$ and $\nu =1$ corresponds to the
Randall--Sundrum type II (R-S II) braneworld scenario 
~\cite{R_S,cline,SMS,Binetruy}, where a co--dimension one brane with positive 
tension $\sigma$ is 
embedded in five--dimensional Anti--de Sitter (${\rm AdS_5}$) space: 
\beqn
\label{RSIIcorrection}
L(\rho) = \sqrt{1 + \frac{\rho}{2 \sigma}}  .
\eeqn

For this case, the scaling potential yielding the 
fixed point solution is given by 
\beqn
\label{rs-pot}
V(\phi) = 2 \sigma Y_c^2 {\rm cosech}^2 
\left(-\frac{\lambda}{2} \frac{\sqrt{8 \pi}}{m_4} \phi \right)
\eeqn
and the time--dependences of the scalar field and scale factor are 
deduced by evaluating the integral (\ref{integrandplus}) for 
$\nu =1$ and substituting the result into Eq. (\ref{plussolution}): 
\begin{eqnarray}
\cosh \left( -\frac{\lambda}{2}\frac{\sqrt{8 \pi}}{m_4} \phi 
\right) =  \left( \frac{2\pi}{3Am_4^2} 
\right)^{1/2} \lambda^2 \Omega_{\phi c} t
\\ 
a (t) = \left[ \frac{2\pi \Omega_{\phi c}^2 \lambda^4}{3Am_4^2} t^2 -1
\right]^{1/(\lambda^2 \Omega_{\phi c})}  .
\label{RSsolution}
\end{eqnarray}

These results for the scaling solutions associated with the R-S II model 
confirm those 
previously obtained using a different method in  Ref.~\cite{Hawkins:2000dq}. 
An important 
feature of our approach is that it shows the 
solution is a fixed point attractor solution. Another important 
feature that emerges is that in the limit where 
the  quadratic energy density term dominates, i.e., when 
$\sqrt{8 \pi} \lambda \phi \ll 2 m_4$, the potential (\ref{rs-pot})
asymptotes to $V \propto \phi^{-2}$, consistent with earlier 
analyses \cite{MMY, Maeda:2000mf}. Similarly, once the energy 
density has decreased so that $L \sim 1$, the value of the scalar 
field becomes large, and  
$V \sim \exp [-\lambda (\sqrt{8 \pi}/ m_4) \phi]$, in 
agreement with general relativistic results \cite{trac,CLW}.

In Figs.~\ref{Fig_r_scaling}-\ref{Fig_r_power1}, we confirm numerically 
how the above potential leads to the 
expected attractor solutions for a model with a barotropic fluid of radiation
($\gamma = 4/3$).

\begin{figure}[h]
\begin{center}
\includegraphics[width=80mm]{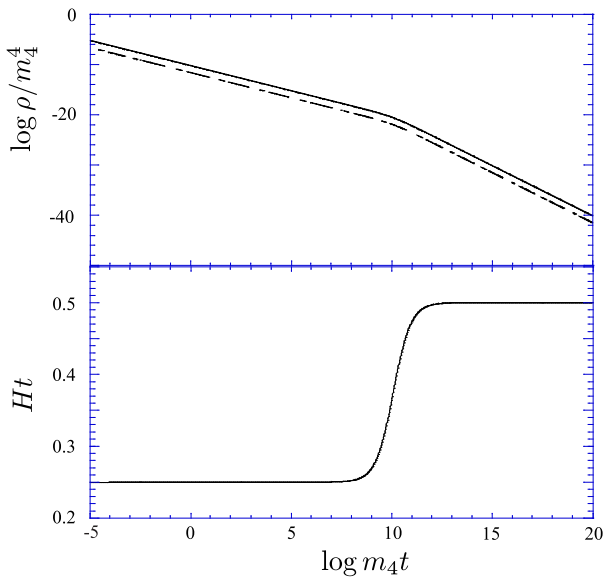}
\caption{Scaling solution for R-S II brane
cosmology including a radiation fluid and a scalar field with  potential
given by Eq.~(\ref{rs-pot}). In order to obtain the scaling solution,
we choose $\lambda=10$ and for simplicity specify  
the energy scale of the brane tension to be  
$\sigma/m_4^4 = 10^{-20}$. In the upper figure, the time evolution
of the energy density
for both  radiation $\rho_{\rm r}$ (solid curve) and the 
scalar field $\rho_\phi$ (dashed curve) are shown. Note that around 
$\log m_4 t \sim 10$ there is a rapid change 
as the quadratic correction becomes 
negligible and the standard cosmological evolution is recovered.
However, the energy density of the scalar field 
mimics that of the radiation fluid throughout the entire evolution, 
i.e., there is scaling behaviour.
In the lower figure, we show the time dependence of the cosmic expansion law.
At early times, before $\log m_4 t \sim 10$, the scale factor grows as 
$a \propto t^{1/4}$, and represents a solution for 
a radiation--dominated universe in a $\rho^2$ dominated cosmology.
After this time, the conventional expansion rate
$a \propto t^{1/2}$ arises. 
}
\label{Fig_r_scaling}
\end{center}
\end{figure}

\begin{figure}[h]
\begin{center}
\includegraphics[width=80mm]{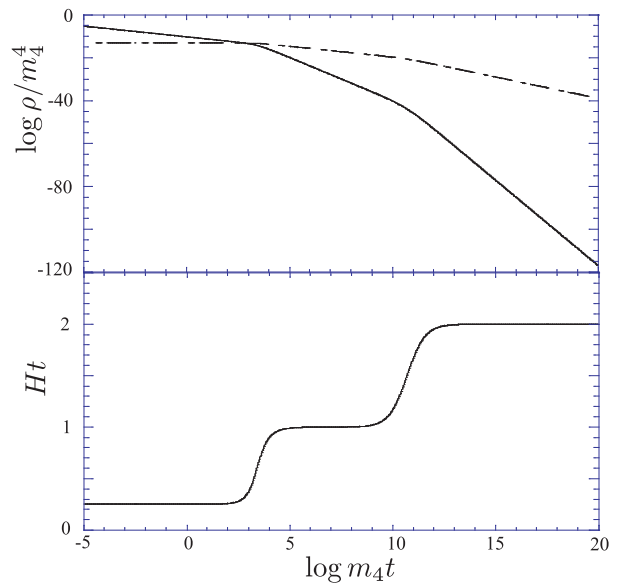}
\caption{As in Fig. (\ref{Fig_r_scaling}), although now 
we specify $\lambda=1$. This leads to a scalar field dominated universe.
In the lower figure, the time dependence of the cosmic 
expansion law is shown. As expected in a universe dominated 
by a scalar field, the scale factor grows as 
$a \propto t$ at early times, corresponding to 
the $\rho^2$ dominated phase, and as 
$a \propto t^2$ at late times when the linear $\rho$ term is important. 
}
\label{Fig_r_power1}
\end{center}
\end{figure}

\subsection{Shtanov-Sahni braneworld cosmology}

The case $A=-1/(2|\sigma |)$ and $\nu =1$ represents a class of 
braneworld inspired cosmologies due to Shtanov and Sahni (S-S)  
\cite{Shtanov:2001pk,Kofinas:2001es,Shtanov:2002mb}. In this 
scenario, a co--dimension one brane with negative tension $\sigma$
is embedded in a five--dimensional 
conformally flat Einstein space, where the signature of the 
fifth dimension is timelike.  
In this model, the deviation from the conventional Friedmann 
cosmology is characterised by
\beqn
L(\rho) = \sqrt{1 - \frac{\rho}{2 |\sigma|}} .
\eeqn
This model has recently been invoked to develop a non--singular 
oscillating universe, where the turning 
points in both the contracting and expanding phases are induced by 
the quadratic correction \cite{Brown:2004cs}.

The S-S braneworld is dual to that of the RS-II scenario 
in the sense discussed 
above. The scaling potential follows directly from Eq. (\ref{g-pot.2}): 
\beqn
\label{ss-pot}
V(\phi) = 2 |\sigma| Y_c^2 {\rm sech}^2 
\left(-\frac{\lambda}{2} \frac{\sqrt{8 \pi}}{m_4} \phi \right)  ,
\eeqn
whereas the time--dependences of the scalar field and scale factor 
follow from Eqs. (\ref{integrandminus}) and 
(\ref{aminussolution}), respectively, after substituting 
for $A=- 1/(2 |\sigma|)$ and $\nu=1$:  
\begin{eqnarray}
\sinh \left( -\frac{\lambda}{2}\frac{\sqrt{8 \pi}}{m_4} \phi 
\right) =  \left( \frac{2\pi}{3|A| m_4^2} 
\right)^{1/2} \lambda^2 \Omega_{\phi c} t
\\ 
a(t) = \left[ 1+ \frac{2\pi \Omega_{\phi c}^2 \lambda^4}{3|A|m_4^2} t^2 
\right]^{1/(\lambda^2\Omega_{\phi c})}   .
\label{S-Ssolution}
\end{eqnarray}

Such a scaling solution is phenomenologically interesting since it 
represents a non--singular bouncing cosmology. The universe 
collapses from infinity $(t\rightarrow -\infty)$ to a finite 
size at $t=0$ and then bounces into an expanding phase. 
The scalar field rolls up the potential during the collapse, 
reaches the maximum of the potential at 
$\phi =0$ at the instant of the bounce, and then 
rolls monotonically down the other side during the expansion
era. 

\subsection{Cardassian cosmology}

In the above classes of models, the modifications to the 
Friedmann equation become significant at high energy scales (early times). 
On the other hand, recent CMB and large--scale structure 
observations indicate that 
the universe is entering a stage of accelerated expansion 
at the present epoch and a number of phenomenological 
models have been developed in an attempt to provide a geometrical 
interpretation of these observations. 
In Cardassian cosmology \cite{Freese:2002sq,Freese:2002gv}, for example, 
the modification 
term in the Friedmann equation is given by Eq. (\ref{L_form}) 
with $A>0$ and $\nu \equiv n $ \cite{Freese:2002sq,Freese:2002gv}:  
\beqn
L(\rho) = \sqrt{1+A \rho^n} 
\eeqn
and the present--day cosmological acceleration can be explained 
even when the energy density is comprised of only 
ordinary matter sources if $n < -1/3$. 
The characteristic feature of this model, therefore, is that
the modification term becomes significant at late times.

Although no scalar field is present in the scenario 
considered in Refs. \cite{Freese:2002sq,Freese:2002gv}, it 
is instructive to show
that the equivalent background cosmology can be obtained from a 
model comprised of a barotropic fluid and a self--interacting scalar field. 
We see immediately from Eq.~(\ref{g-pot.1}) that 
the corresponding potential which provides 
the fixed point attractor solution is given by
\beqn
\label{ca-pot}
V(\phi) = Y_c^2 A^{-\frac{1}{n}} {\rm cosech}^{\frac{2}{n}}
\left(- \frac{\lambda n}{2} \frac{\sqrt{8 \pi}}{m_4} \phi \right) 
\eeqn
and in Fig.~\ref{Fig_c_scaling} we 
demonstrate numerically how the above potential leads to the expected 
attractor solution. Although such a homogeneous solution 
is indistinguishable from the purely perfect fluid 
background, it is possible 
that the presence of a scalar field may modify the evolution 
of perturbations and the clustering properties 
of matter. In principle, this could  
result in potentially observational signatures \cite{Bartolo:2003ad}
and a scaling solution of this 
type provides a framework for quantitatively investigating the 
evolution of perturbations in these models. 

\begin{figure}[h]
\begin{center}
\includegraphics[width=80mm]{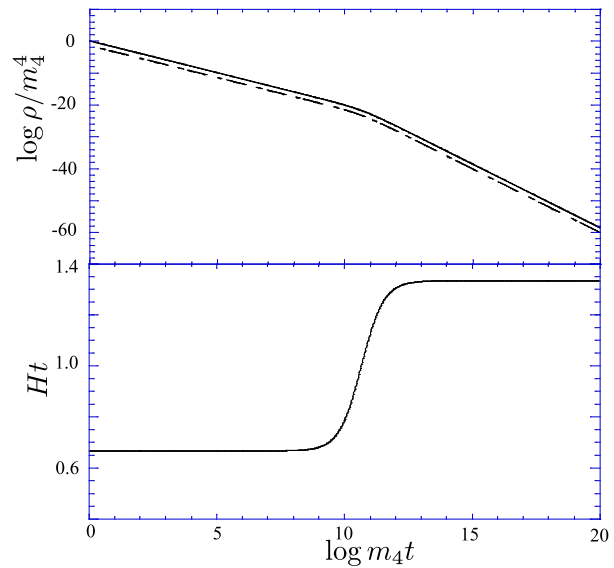}
\caption{Scaling solution for a Cardassian--type 
cosmology including a matter fluid source and a scalar field 
with a potential given by Eq.~(\ref{ca-pot}). We specify 
$n=-0.5$  and choose $\lambda=10$ in order to obtain the scaling solution. 
For simplicity, we set the energy scale where modifications 
to the standard scenario become significant to be $A/m_4^2 = 10^{-10}$. 
In the upper figure, the time evolution
of the energy density of matter $\rho_{\rm m}$ (solid curve)
and that of the scalar field $\rho_{\phi}$ are shown.
When $\log m_4 t \sim 10$, the time--dependence of 
the energy density changes as the late-time modification term 
becomes significant. In the lower figure, we show the time dependence of the
cosmic expansion law.
For $\log m_4 t \sim 10$, we see that $a \propto t^{2/3}$,  
corresponding to the conventional matter dominated universe,
whereas for  $\log m_4 t  > 10$, the correction term leads to an accelerating 
universe, $a \propto t^{4/3}$. 
}
\label{Fig_c_scaling}
\end{center}
\end{figure}

Finally, before concluding this Section, we illustrate the 
duality transformation that relates the Cardassian cosmologies 
with the S-S braneworld. Denoting the former with a subscript 
`$+$' and the latter by `$-$', we may substitute 
the form for $b_+ (\phi )$ in 
the Cardassian scenario, as given by Eq. (\ref{plussolution}), into 
Eq. (\ref{energydual}) to deduce that 
\begin{equation}
\label{carddualpot}
V_-(\phi ) \propto {\rm sech}^{-2p\Omega_{\phi c}^2 \lambda_-^2/n}
\left( - \frac{\lambda_+ n}{2} \frac{\sqrt{8\pi}}{m_4} \phi
\right)   .
\end{equation} 
This reduces to the S-S scaling potential Eq.~(\ref{g-pot.2}) (with 
$\nu =1$ and $\lambda = \lambda_-$) when $p \Omega_{\phi c}^2\lambda_-^2 = 
-n$ and $\lambda_+ = \lambda_-/n$. Moreover, in this 
case, it can be verified that the dual correction function 
satisfying Eq. (\ref{necessary}) reduces to Eq. (\ref{minusL}) 
with $\nu= 1$ and $\lambda = \lambda_-$. 

We now proceed in following section to determine the scaling 
solutions in a further braneworld 
scenario, where the corrections to the Friedmann equation 
become significant at late times. 

\section{Dvali-Gabadadze-Porrati braneworld cosmology}

The Dvali--Gabadadze--Porrati (DGP) braneworld scenario 
\cite{Dvali:2000hr,Deffayet:2000uy} corresponds to a 3--brane embedded
in flat five--dimensional Minkowski spacetime, where a Ricci scalar term
is included in the brane action. All energy--momentum is confined to the 
brane since the bulk is empty. The modified Friedmann equation for the 
DGP model is given by \cite{Deffayet:2000uy}
\begin{equation}
\label{DGPFriedmann}
H^2 \pm \frac{H}{r_0} = \frac{8\pi}{3m^2_4} \rho  ,
\end{equation}
where $r_0 \equiv m_4^2/(2m_5^3)$ and $m_5$ is the five--dimensional 
Planck scale. In the DGP model, gravity behaves as  
four--dimensional Einstein gravity at short scales, 
whereas it propagates into the bulk 
at large scales. This induces corrections to the 
standard Friedmann equation at low energies and 
the parameter $r_0$ determines the scale at which these corrections
become important. There are two 
inequivalent ways of embedding the brane in the bulk and 
this is reflected in the different choices of sign in Eq. 
(\ref{DGPFriedmann}). In this section, 
we refer to these as the $(+)$ and $(-)$ branches, respectively. 
In subsequent expressions, where different signs may be taken,   
the upper case corresponds to the $(+)$ branch. 

It proves convenient to express the Friedmann equation
(\ref{DGPFriedmann}) in the equivalent form 
\begin{equation}
\label{DGPFriedmann1}
H= \frac{1}{2r_0} \left[ \mp 1+ \sqrt{1+ B\rho} \right]  ,
\end{equation}
where 
\begin{equation}
\label{defB}
B \equiv \frac{32\pi r_0^2}{3m_4^2}   .
\end{equation}
For the $(-)$ branch, the late--time attractor is de Sitter (exponential) 
expansion for any decreasing energy density \cite{Deffayet:2000uy}. 
For the $(+)$ branch, on the other hand, expanding Eq. (\ref{DGPFriedmann1}) 
as a Taylor series to lowest--order implies that $H \approx 8 \pi r_0
\rho /(3m_4^2)$. Modulo a rescaling of the 
four--dimensional Planck mass, this corresponds formally to the
high--energy limit $(\rho \gg 2\sigma )$ of the R-S II braneworld 
(\ref{RSIIcorrection}). Consequently, the early--time 
analysis of the latter model performed in Ref. \cite{Mizuno:2004xj} 
is directly applicable to the late--time 
behaviour of this branch of the DGP model. In particular, 
we may conclude immediately that the potential driving the scaling solution 
in this limit is the inverse power--law potential $V \propto \phi^{-2}$. 

A direct comparison between Eqs. (\ref{b_eq}) 
and (\ref{DGPFriedmann1}) implies that the Friedmann 
correction function is given by 
\begin{equation}
\label{DGPcorrect}
L = \frac{1}{\sqrt{B\rho}} \left[ \mp 1 + \sqrt{1+B \rho} \right]  .
\end{equation}
In order to derive the scaling solutions, we 
define a new variable, $\theta$:  
\begin{equation}
\label{deftheta}
\rho \equiv \frac{1}{B} \sinh^2 \theta . 
\end{equation}
Substituting Eq. (\ref{DGPcorrect}) into Eq. (\ref{dif_V_2}) 
then implies that the solution represents a scaling solution if 
\begin{equation}
\label{DGPscaling}
-\frac{\sqrt{8\pi} \lambda}{m_4} \phi = 
2 \int d \theta \, \frac{\cosh \theta}{\cosh \theta \mp 1}  
\end{equation}
and the integral (\ref{DGPscaling}) may be evaluated to yield 
the form of the potential: 
\begin{equation}
\frac{\sqrt{2\pi} \lambda}{m_4} \phi =
\frac{\sqrt{B\rho}}{\sqrt{1+B\rho} \mp 1}
- {\rm sinh}^{-1} \sqrt{B \rho}  .
\end{equation}

The corresponding 
time dependence of the scaling solution can also be determined. 
In terms of the variable (\ref{deftheta}), the Friedmann equation 
(\ref{DGPFriedmann1}) simplifies to 
\begin{equation}
\label{Friedmann+}
H(\theta ) = \frac{1}{r_0} \sinh^2 \frac{\theta}{2}
\end{equation}
for the $(+)$ branch, and 
\begin{equation}
\label{Friedmann-}
H(\theta ) = \frac{1}{r_0} \cosh^2 \frac{\theta}{2}
\end{equation}
for the $(-)$ branch. 
Recalling that $\dot{\phi}^2 /\rho = 2 X_c^2$ and 
$\rho_{\phi} = \Omega_{\phi c} \rho$ for the scaling solution, 
it follows that the scalar field equation (\ref{rhoeom}) 
transforms to 
\begin{equation} 
\label{DGPtdep}
t = - \frac{\Omega_{\phi c}}{3X_c^2} \int d \theta \, 
\frac{{\rm cotanh} \, \theta}{H(\theta ) }
\end{equation}
after substitution of Eq. (\ref{deftheta}). 
Substituting Eqs. (\ref{Friedmann+}) and (\ref{Friedmann-}) for the 
$(+)$ and $(-)$ branches, respectively, and evaluating 
the integral (\ref{DGPtdep}) then implies that 
\begin{equation}
\label{DGPtimesol}
\frac{3X_c^2}{\Omega_{\phi c}r_0} t = {\rm cotanh}^{-1} 
\sqrt{1+B\rho} + \frac{1}{\sqrt{1+B\rho} \mp 1}   .
\end{equation}

\section{Summary}

In this paper we have brought together a number of recent 
approaches to cosmology which 
involve modifications of the Friedmann equation. By introducing the 
general function $L(\rho)$ as the way of parametrising 
the modification, we have been able to establish the conditions under 
which the new system enters scaling solutions. Considering the case 
where the energy density is comprised of a scalar field and background 
barotropic fluid, we have obtained the general 
relationship that would have to be satisfied between 
the  evolving scalar field and  $L(\rho)$. In particular, we have 
obtained the corresponding potential $V(\phi)$ which leads to scaling 
solutions and, for a rather general class of functions of $L(\rho)$, we 
have shown that there exist dual solutions which also exhibit similar 
scaling behaviour. This has allowed us to relate solutions which would 
otherwise appear quite distinct, including those involving collapsing 
and expanding cosmologies. 

Moreover, the duality can directly relate singular and non--singular 
cosmologies. To illustrate this property,  
consider a particular scaling solution $a_+(\phi )$ that is 
singular in the sense that the scale 
factor vanishes at $a_+(0) =0$. (The value of the scalar 
field can be chosen to be $\phi =0$ without loss of generality). 
Suppose, however, that the logarithmic  derivative of the scale factor 
with respect to the field is non--zero at this 
point, $d \ln a_+/d\phi |_0 \ne 0$, and furthermore, that  
the scale factor is a monotonic function with a finite  
first derivative for all physical (non--zero) values of the field. These 
properties are satisfied, for example, in 
the R-S II and Cardassian models.  

The qualitative behaviour of the dual solution, $a_-(\phi )$, is 
then determined from Eq. (\ref{dualscale}). If we define a new parameter 
$\epsilon (\phi )\equiv [d \ln a /d\phi ]^{-1}$, the value of the scale 
factor is simply given by the area under the curve $\epsilon_+ (\phi )$,  
where the field evolves from zero to some value $\phi$. (We are assuming 
implicitly that $\dot{\phi} >0$ and $p<0$ again 
without loss of generality). However, 
due to the exponential nature of Eq. (\ref{dualscale}) the initial 
value of the dual scale factor is {\em non--zero} and 
this results in a non--singular background. 

On the other hand, the time reversal of the seed solution $a_+(\phi )$ would 
result in a collapsing dual model where the limits in 
the integral (\ref{dualscale}) are taken from $\phi$ to 
zero. Consequently, the dual solution can be analytically 
continued through $\phi =0$ into a contracting phase. In this sense, 
therefore, any singular (expanding) scaling solution satisfying 
the above (very weak) conditions can generate 
a non--singular bouncing cosmology, where 
the latter is associated with a combination of the seed solution 
and its time reversal. For fixed values of $\{ p, 
\lambda_{\pm} \}$, the collapsing 
phase of the bouncing solution will be unstable if the 
expanding phase is stable, and vice--versa. In principle, 
however, different seed solutions may be employed to generate 
distinct and stable collapsing and expanding branches 
that can be smoothly joined at the bounce. 

This opens up the possibility that such dualities will 
allow us to relate singular cosmologies to 
non-singular bouncing cosmologies, a topic presently of considerable  
interest in cosmology. 

Finally, we have argued that the type of correction given by Eq. (\ref{L_form}) 
arises in a number of particle physics motivated models. Further 
examples arise in the limit where $A\rho^{\nu} \gg 1$. In particular, the case 
$\nu =-1/3$ corresponds to the high--energy 
limit of the Gauss--Bonnet braneworld \cite{gb}. In this model, 
the R-S II scenario is generalized to 
include a Gauss--Bonnet combination of curvature
invariants in the five--dimensional bulk action. 
More generally, effective Friedmann equations of the 
form $H^2 \propto \rho^{\nu}$, where $\nu$ is arbitrary \cite{chung}, 
can arise in models based on Ho\v{r}ava--Witten 
theory compactified on a Calabi--Yau three--fold \cite{H_W}. 
Generalized scaling solutions driven by corrections of this form 
were recently investigated for a variety of scalar field models
\cite{Tsujikawa:2004dp}.

Future directions involving the use of the duality properties of 
these models would include an extension of our analysis to negative potentials, 
thereby allowing us to link these classes of solutions with those arising 
in the cyclic/ekpyrotic scenario \cite{Steinhardt:2001st}.  
On the other hand, as we have seen, the more general corrections proposed 
in \cite{Dvali:2000hr,Deffayet:2000uy} that arise due to modifications
of gravity on large scales can lead to an explanation of  
the present cosmic acceleration without introducing dark 
energy \cite{Lue:2004za}. It would be interesting to investigate the
impact that the duality transformations we have described 
have on such a scenario. 

\section*{Acknowledgments}
EJC would like to thank the Aspen Center for Physics for their 
hospitality during the time part of this paper was completed.
SM~would like to thank Kei-ichi Maeda for continuous
encouragement and is grateful to the University of Sussex 
for their hospitality during a period when this work 
was initiated.

\end{document}